\documentclass[12pt,prb,superscriptaddress,showpacs,preprintnumbers,amsmath,amssymb]
{revtex4}

\usepackage{graphicx} % Include figure files
\usepackage{dcolumn} % Align table columns on decimal point

\def\eps{\epsilon}
\def\gam{\gamma}

\def\lam{\lambda}

\def\om{\omega}

\def\sg{\sigma}
\def\Sg{\Sigma}

\def\bd{{\bf d}}

\def\bk{{\bf k}}

\def\bp{{\bf p}}
\def\bq{{\bf q}}

\def\bv{{\bf v}}

\def\b0{{\bf 0}}

\def\cO{{\cal O}}

\def\Im{{\rm Im}}

\def\sgn{{\rm sgn}}
\def\det{{\rm det}}

\begin{document}

\title{Singular order parameter interaction at nematic quantum critical
 point in two dimensional electron systems}

\author{Stephan C.~Thier}
\thanks{Present address: Institut f\"ur Physik, 
Johannes-Gutenberg-Universit\"at, 55099 Mainz, Germany}
\affiliation{Universit\"at Stuttgart, Fachbereich Physik,
 D-70550 Stuttgart, Germany}
\affiliation{Max-Planck-Institute for Solid State Research,
 D-70569 Stuttgart, Germany}
\author{Walter Metzner}
\affiliation{Max-Planck-Institute for Solid State Research,
 D-70569 Stuttgart, Germany}

\date{\today}

\begin{abstract}
We analyze the infrared behavior of effective $N$-point 
interactions between order parameter fluctuations for nematic
and other quantum critical electron systems with a scalar 
order parameter in two dimensions.
The interactions exhibit a singular momentum and energy
dependence and thus cannot be represented by local vertices. 
They diverge for all $N \geq 4$ in a collinear low-energy limit, 
where energy variables scale to zero faster than momenta, 
and momenta become increasingly collinear.
The degree of divergence is not reduced by any cancellations
and renders all $N$-point interactions marginal.
A truncation of the order parameter action at quartic or any 
other finite order is therefore not justified.
The same conclusion can be drawn for the effective action
describing fermions coupled to a $U(1)$ gauge field in two
dimensions.
\end{abstract}

\maketitle

%%% Introduction %%%%%%%%%%%%%%%%%%%%%%%%%%%%%%%%%%%%%%%%%%%%%%%%%%%%%%%%%%

\section{Introduction}

Numerous interacting electron systems undergo a quantum phase transition
\cite{sachdev99} between ground states with different symmetries, which 
can be tuned by a non-thermal control parameter such as doping, pressure, 
or a magnetic field.
In the vicinity of a continuous transition electronic excitations are 
strongly scattered by critical order parameter fluctuations, such that 
Fermi liquid theory breaks down.\cite{vojta03,loehneysen07}
Quantum critical fluctuations near a quantum critical point (QCP) are
therefore frequently invoked as a mechanism for non-Fermi liquid behavior
in strongly correlated electron compounds.

Quantum criticality in metallic electron systems is traditionally 
described by an effective order parameter theory which was pioneered 
by Hertz \cite{hertz76} and extended to finite temperatures by Millis. 
\cite{millis93}
In that approach an order parameter field $\phi$ is introduced via
a Hubbard-Stratonovich decoupling of the electron-electron interaction,
and the electronic variables are subsequently integrated out.
The resulting effective action $S[\phi]$ for the order parameter is
truncated at quartic order and analyzed by standard scaling techniques.

However, several studies revealed that the Hertz-Millis approach may
fail, especially in low-dimensional systems.\cite{belitz05,loehneysen07}
Since electronic excitations in a metal are gapless, integrating out
the electrons may lead to singular interactions between the order 
parameter fluctuations which cannot be approximated by a local
quartic term.
The nature of the problem and essential aspects of its solution were
presented first for disordered ferromagnets by Kirkpatrick and 
Belitz.\cite{kirkpatrick96}
For clean ferromagnets, Belitz et al.\cite{belitz97} showed that
Hertz-Millis theory breaks down, and no continuous quantum phase
transition can exist, in any dimension $d \leq 3$; the transition is
generically of first order.\cite{belitz99}
The Hertz-Millis approach was also shown to be invalid for the 
quantum antiferromagnetic transition in two dimensions.
\cite{abanov03,abanov04,metlitski10sdw}
In that case a continuous transition survives, but the QCP becomes
non-Gaussian.

In this article we analyze the validity of the Hertz-Millis approach
to quantum criticality for two-dimensional systems with singular 
{\em forward scattering} of electrons, in particular systems exhibiting
a quantum phase transition driven by forward scattering in the charge
channel.
The most prominent such transition is the electronic {\em nematic}, 
in which an orientation symmetry is spontaneously broken, while 
translation and spin-rotation invariance remain unaffected.\cite{fradkin10}
The problem of quantum critical points with singular forward scattering
is formally similar to the problem of non-relativistic fermions coupled
to a $U(1)$ gauge field, which was studied intensively in the 1990s.
\cite{metzner98,lee06}

Perturbation theory for the electronic self-energy at the nematic QCP 
yields a non-Fermi liquid contribution proportional to $\om^{2/3}$
already at the lowest order in a loop expansion.\cite{oganesyan01,metzner03}
The same behavior was found earlier for fermions coupled to a $U(1)$ 
gauge field.\cite{lee89}
It was commonly believed that the power-law with an exponent $\frac{2}{3}$ 
is not modified by higher order contributions.
Furthermore, calculations in the gauge field context suggested that 
the simple form of the (bosonic) fluctuation propagator obtained in 
lowest order (RPA) remains unaffected by higher order terms.\cite{kim94}
In a remarkable recent paper Metlitski and Sachdev \cite{metlitski10nem}
formulated a scaling theory of the nematic QCP and related problems, 
treating the electrons and order parameter fluctuations on equal 
footing. In a renormalization group calculation they found a
logarithmic divergence at three-loop order pointing at a correction 
of the $\om^{2/3}$ law for the electronic self-energy.
However, no qualitative correction was found for the fluctuation 
propagator, up to three-loop order.\cite{footnote1,footnote2}
This is in stark contrast to the case of an antiferromagnetic QCP 
in two dimensions, where the fluctuation propagator is substantially 
renormalized compared to the RPA form.\cite{abanov03,metlitski10sdw}
A clarification of the properties of the nematic QCP beyond three-loop 
order is still lacking. 

The robustness of the fluctuation propagator at the nematic QCP seems 
to indicate that interactions of the order parameter fluctuations are 
irrelevant such that the QCP is Gaussian, in agreement with the 
expectations from Hertz-Millis theory.
It is therefore worthwhile to analyze the interaction terms in the
effective action $S[\phi]$ obtained after integrating out the 
electrons. The $N$-point interactions are given by fermionic loops 
with $N$ vertices. To obtain the scaling behavior of such loops is
non-trivial, because the most naive power-counting is easily 
invalidated by cancellations.\cite{neumayr98,kopper01} 
In this paper we compute the exact scaling behavior of the 
$N$-point interactions for the nematic QCP and related systems. 
We find that the interactions are {\em marginal} and {\em non-local} 
for all $N \geq 3$.
Hence, replacing them by a local $\phi^4$ interaction is not
justified.

The paper is structured as follows.
In Sec.~II we introduce the effective actions to be analyzed, 
and we define the $N$-point loops describing the interaction 
terms.
In Sec.~III we explain the special role of fluctuations with
collinear momenta, which motivates the definition of the
collinear low-energy scaling limit.
Secs.~IV-VI are dedicated to the analysis of the $N$-point 
loops.
After reviewing exact formulae from the literature (Sec.~IV), 
we derive explicit expressions for $N$-point loops in the 
collinear low-energy scaling limit.
Using the scaling behavior of these loops, we perform the 
power counting of $N$-point order parameter interactions in 
Sec.~VII. We finally conclude in Sec.~VIII.

%%% Effective action and fermion loops %%%%%%%%%%%%%%%%%%%%%%%%%%%%%

\section{Effective action and N-point loops}

We consider an interacting Fermi system which undergoes a continuous
quantum phase transition with a scalar order parameter of the form
\begin{equation} \label{orderpar}
 O = \sum_{\sg} \int \frac{d^2 k}{(2\pi)^2} \, d_{\sg}(\bk) 
 c_{\sg}^{\dag}(\bk) c_{\sg}^{\phantom\dag}(\bk) \, ,
\end{equation}
where $c_{\sg}^{\dag}(\bk)$ and $c_{\sg}^{\phantom\dag}(\bk)$
are the usual fermionic creation and annihilation operators.
For a charge nematic \cite{fradkin10} the form factor $d_{\sg}(\bk)$ 
is spin symmetric and has a $\bk$-dependence with $d$-wave 
symmetry, such as $d_{\sg}(\bk) = \cos k_x - \cos k_y$. 
A spin-antisymmetric form factor may describe an Ising ferromagnet
or an Ising spin nematic.

Decoupling the fermionic interaction by introducing an order
parameter field $\phi$ via a Hubbard-Stratonovich transformation, 
and integrating out the fermionic variables,\cite{hertz76} one
obtains an effective action
\begin{eqnarray} \label{action}
 S[\phi] &=& \frac{1}{2} \int_q g^{-1} \phi(q) \phi(-q)
 \nonumber \\ 
 &+& \sum_{N=2}^{\infty} \frac{(-1)^N}{N} \int_{q_1,\dots,q_N} 
 \delta(q_1 + \dots + q_N) \, \Pi_{d,N}(q_1,\dots,q_N) \,
 \phi(q_1) \dots \phi(q_N) \, ,
\end{eqnarray}
where $g > 0$ is the fermionic coupling constant, and
\begin{eqnarray} \label{Pi_dN}
 \Pi_{d,N}(q_1,\dots,q_N) = \sum_{\sg} \int_k \, \prod_{j=1}^N 
 \left[ d_{\sg}(\bk - \bp_j - \bq_j/2) \, 
 G_0(k - p_j) \right] \, .
\end{eqnarray}
Here and in the following we use 3-vectors collecting imaginary 
frequency and two-dimensional momentum variables, 
for example $k = (k_0,\bk)$, and $\int_k$ as a short-hand 
notation for $\int \frac{dk_0}{2\pi} \frac{d^2k}{(2\pi)^2}$.
The variables $p_j$ and $q_j$ are related by
\begin{eqnarray} \label{pq}
 q_j &=& p_{j+1} - p_j \quad \mbox{for} \quad j = 1,\dots,N-1 
 \nonumber \\
 q_N &=& p_1 - p_N \, .
\end{eqnarray}
Note that $q_1 + \dots + q_N = 0$ due to energy and momentum
conservation.
The bare propagator has the form
$G_0(k) = \left[ik_0 - \eps(\bk) + \mu \right]^{-1}$, where 
$\eps(\bk)$ is the dispersion relation of the non-interacting 
particles.
$\Pi_{d,N}(q_1,\dots,q_N)$ can be represented graphically as a 
fermion loop with $N$ lines corresponding to $G_0$ and $N$ vertices
with form factors $d_{\sg}(\bk)$, as shown in Fig.~1.
For spin-antisymmetric form factors, $\Pi_{d,N}$ vanishes for 
odd $N$.
\begin{figure}[htb]
\centering
\includegraphics[width=0.35\textwidth]{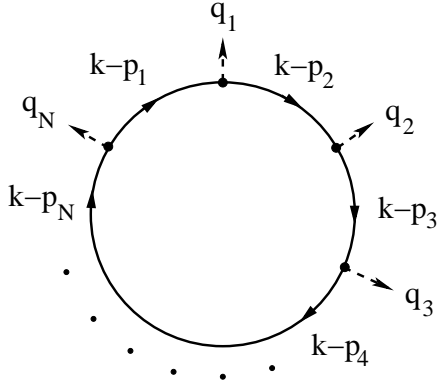}
\caption{Graphical representation of $\Pi_{d,N}$ with momentum
 variables as in Eq.~(\ref{Pi_dN}).}
\end{figure}

For fermions coupled to a $U(1)$ gauge field, integrating out the 
fermions leads to a similar effective action $S[\phi]$, where $\phi$ 
is the transverse component of the gauge field (in Coulomb gauge).
The bosonic $N$-point functions are then essentially given by a loop
with transverse current vertices
\begin{eqnarray} \label{Pi_tN}
 \Pi_{t,N}(q_1,\dots,q_N) = 2 \int_k \, \prod_{j=1}^N 
 \left[ \hat\bq_{j\perp} \cdot \bv(\bk - \bp_j - \bq_j/2) \, 
 G_0(k - p_j) \right] \, ,
\end{eqnarray}
where $\bv(\bk) = \nabla\eps(\bk)$, and $\hat\bq_{\perp}$ is the unit
vector obtained by rotating $\hat\bq = \bq/|\bq|$ by $\pi/2$, that is,
$\hat\bq_{\perp} = (-\hat q_y, \hat q_x)$.
In addition there are contributions from the ``diamagnetic'' term
(of the form $\phi^2\bar\psi\psi$) in the underlying fermionic action, 
which are however less singular, since they involve a smaller 
number of propagators (less than $N$).

The $N$-point contribution to $S[\phi]$ in Eq.~(\ref{action}) is
symmetric under any permutation of $q_1,\dots,q_N$. Hence,
one can replace $\Pi_{d,N}$ by the symmetrized $N$-point loop
\begin{equation} \label{Pi_dN^S}
 \Pi_{d,N}^S(q_1,\dots,q_N) = \frac{1}{N!} \sum_{P}
 \Pi_{d,N}(q_{P1},\dots,q_{PN}) \, ,
\end{equation}
where the sum collects all permutations $P$ of $1,\dots,N$.
Non-symmetric contributions to $\Pi_{d,N}$ do not contibute to the
integral in Eq.~(\ref{action}).
For the gauge field problem one defines $\Pi_{t,N}^S$ analogously.
Substantial cancellations may occur in the sum over permutations.
\cite{neumayr98,kopper01}

%%% Collinear low-energy limit %%%%%%%%%%%%%%%%%%%%%%%%%%%%%%%%%%%%%%

\section{Collinear low-energy limit}

The quadratic part of $S[\phi]$ is determined by a constant and the
2-point loop or ``bubble''
\begin{equation} \label{bubble}
 \Pi_d(q) = \Pi_{d,2}(q,-q) = \sum_{\sg} \int_k d_{\sg}^2(\bk) \,
 G_0(k-q/2) G_0(k+q/2) \, .
\end{equation}
For small $\bq$ and small $|q_0|/|\bq|$, it has an expansion of
the form \cite{hertz76,dellanna06}
\begin{equation}
 \Pi_d(q) = - N_d + \chi_d \bq^2 + \gam_d \frac{|q_0|}{|\bq|}
 + \dots \, ,
\end{equation}
where 
$N_d = \sum_{\sg} \int \frac{d^2k}{(2\pi)^2} \, d_{\sg}^2(\bk) 
 \delta[\eps(\bk) - \mu]$ 
is a weighted density of states, and $\chi_d$ and $\gam_d$ are
two other constants.
At the quantum critical point one has $g^{-1} - N_d = 0$ such 
that the quadratic part of the action vanishes for $\bq \to 0$ 
and $q_0/|\bq| \to 0$.
For the gauge problem, the constant $N_t$ from the static limit
of $\Pi_t(q)$ cancels generically against a tadpole contribution,
\cite{metzner98,lee06} such that the theory is always critical.
In both cases, the Gaussian part of the action has thus the 
asymptotic form
\begin{equation} \label{S0}
 S_0[\phi] = \frac{1}{2} \int_q 
 \Big( \chi \bq^2 + \gam \frac{|q_0|}{|\bq|} \Big) \,
 \phi(q) \phi(-q) \, ,
\end{equation}
corresponding to a bare propagator
\begin{equation} \label{D0}
 D_0(q) = \frac{1}{\chi \bq^2 + \gam \frac{|q_0|}{|\bq|}} \, .
\end{equation}

$D_0(q)$ diverges in the limit $\bq \to 0$ and $q_0/|\bq| \to 0$. 
The two terms in the denominator of $D_0(q)$ vanish at the 
same pace for $\bq \to 0 \,$ if $q_0 \propto |\bq|^3$. 
Therefore, the bare dynamical scaling exponent is $z=3$.
To assess the size of the interaction terms in $S[\phi]$ 
one thus has to study the $N$-point loops in a low-energy 
limit with $q_{j0} \propto |\bq_j|^3$.
Naively one would expect that this corresponds to the 
{\em static} limit, where $q_{j0} \to 0$ before $\bq_j \to 0$.
Hertz and Klenin \cite{hertz74} showed that an $N$-point 
loop converges to the $(N-2)$-th derivative of the density 
of states with respect to the Fermi energy in the static
limit.
In our case, with a form factor $d_{\sg}(\bk)$, their 
result generalizes to
\begin{equation} \label{statlim}
 \lim_{\bq_j \to 0} \lim_{q_{j0} \to 0} 
 \Pi_{d,N}(q_1,\dots,q_N) = 
 \frac{(-1)^{N-1}}{(N-1)!} \, 
 \frac{\partial^{N-2}}{\partial\mu^{N-2}}
 \sum_{\sg} \int \frac{d^2k}{(2\pi)^2} \,
 d_{\sg}^N(\bk) \, \delta[\eps(\bk) - \mu] \, .
\end{equation}
Except for special cases where the chemical potential lies 
at a van Hove singularity, this expression is {\em finite}.
Note that the right hand side of Eq.~(\ref{statlim}) is 
independent of $q_1,\dots,q_N$ and hence already symmetrized.
Approximating the bosonic $N$-point interactions by finite 
local interactions thus seems adequate. 
Standard power counting then implies that all interactions
with $N \geq 4$ are increasingly (with higher $N$) irrelevant.
Hence, the Hertz-Millis truncation seems justified for QCPs 
with singular forward scattering, even in two dimensions.
The static limit of the $3$-point loop and all other 
$N$-point loops with odd $N$ as given by Eq.~(\ref{statlim})
vanishes even in the case of a charge nematic, due to the
antisymmetry of $d_{\sg}(\bk)$ under $\pi/2$ rotations of
$\bk$.

One arrives at a similar conclusion for the gauge field 
problem. In that case the static limit of $\Pi_{t,N}$ is
also generically finite. 
Odd $N$-point interactions vanish due to the antisymmetry
of $\bv(\bk)$ and even $N$-point interactions appear to be
irrelevant for any $N \geq 4$.
A Gaussian fixed point thus seems natural.

However, there is a flaw in the above argument.
Eq.~(\ref{statlim}) has been derived by setting $q_{j0} = 0$
before the momenta $\bq_j$ tend to zero.
It is not guaranteed that this captures the low-energy
limit $\bq_j \to 0$ and $q_{j0}/|\bq_j| \to 0$ in general.
Indeed, a simple estimate indicates that the $N$-point loop
is of order $q_{j0}/|\bq_j|^{N-1}$ for small non-collinear 
momenta $\bq_j$ and small $q_{j0}/|\bq_j|$.\cite{metlitski10nem}
Although this behavior is increasingly singular for larger
$N$, the corresponding order parameter interactions remain
irrelevant, since the singularity is not strong enough.
\cite{metlitski10nem}
However,
an even stronger singularity is obtained in a special 
low-energy limit in which the momenta $\bq_1,\dots,\bq_N$ 
become {\em collinear}.
The crucial role of coupled fluctuations with collinear 
momenta was highlighted very clearly by Metlitski and 
Sachdev.\cite{metlitski10nem}
In perturbative one-loop calculations of the fermionic 
self-energy $\Sg(\bk_F,\om)$ at a certain point $\bk_F$ on the 
Fermi surface, it was found already some time ago that the 
dominant contributions involve only fermionic states in the 
momentum region near $\bk_F$ and $-\bk_F$, with momentum 
transfers $\bq$ almost {\em tangential}\/ to the Fermi surface 
in those points. See, for example, Ref.~\onlinecite{lee89} 
for an early calculation in the gauge field context, and 
Ref.~\onlinecite{metzner03} for a corresponding calculation 
at the nematic QCP.
This remains true for higher order contributions,
\cite{metlitski10nem} so that all fermionic momenta are
close to $\bk_F$ and $-\bk_F$ and momentum transfers are
almost tangential to the Fermi surface in these points, which
implies that they are mutually almost collinear.

Choosing a coordinate system in momentum space in such a way 
that the normal vector to the Fermi surface at $\bk_F$ points 
in $x$-direction, the proper scaling limit describing the 
low-energy behavior is given by $k_0 \mapsto \lam^3 k_0$, 
$k_x \mapsto \lam^2 k_x$, and $k_y \mapsto \lam k_y$ with 
$\lam \to 0$, 
where $(k_x,k_y)$ is measured relative to $\bk_F$.
\cite{polchinski94,altshuler94,metlitski10nem}
For the momentum and energy transfers $q_j$ this implies
\begin{equation} \label{colllimit}
 q_{j0} \mapsto \lam^3 q_{j0} \, , \quad
 q_{jx} \mapsto \lam^2 q_{jx} \, , \quad
 q_{jy} \mapsto \lam   q_{jy}
\end{equation}
with $\lam \to 0$. In this {\em collinear low-energy limit} the 
momentum transfers $\bq_j$ become increasingly collinear
(pointing in $y$-direction).
The behavior of the $N$-point interactions given by 
$\Pi_{d,N}(q_1,\dots,q_N)$ and $\Pi_{t,N}(q_1,\dots,q_N)$ 
in the collinear low-energy limit has not yet been studied 
systematically. In particular, it has not yet been analyzed 
whether cancellations suppress their value below the naive 
power counting estimate.
To clarify this issue is the main purpose of our article.

As mentioned above, the dominant contributions are
due to momenta $\bk$ close to those points $\pm \bk_F$ on
the Fermi surface at which the momentum transfers $\bq_j$
are tangent. 
In the definining equation (\ref{Pi_dN}) for $\Pi_{d,N}$
we can therefore replace the form factors $d_{\sg}(\bk)$
by $d_{\sg}(\pm\bk_F)$. 
Assuming $d_{\sg}(-\bk_F) = d_{\sg}(\bk_F)$, which is 
satisfied in all cases of interest, we then obtain
\begin{equation} \label{Pi_dN_N}
 \Pi_{d,N}(q_1,\dots,q_N) \to
 \sum_{\sg} d_{\sg}^N(\bk_F) \, \Pi_N(q_1,\dots,q_N)
\end{equation}
in the collinear low-energy limit, where
\begin{equation} \label{Pi_N}
 \Pi_N(q_1,\dots,q_N) = 
 I_N(p_1,\dots,p_N) =
 \int_k \prod_{j=1}^N G_0(k - p_j)
\end{equation}
is the $N$-point loop for spinless fermions with unit
vertices.
Furthermore, the dispersion $\eps(\bk)$ enters only via the 
Fermi velocity $v_F$ and the Fermi surface curvature in 
$\pm\bk_F$. Both are assumed to be finite, which is the
generic case.
We choose units such that $v_F$ and the curvature radius
$\rho_F$ are both one, and we realize these parameters by
using a simple parabolic dispersion relation 
$\eps(\bk) = \bk^2/2$ and setting $k_F = 1$.
Relating $\Pi_{d,N}$ to $\Pi_N$ with a quadratic dispersion
enables us to exploit exact results for $\Pi_N$ which are 
already available (see below).
For the gauge field problem the vertices are antisymmetric
under reflections, since $\bv(-\bk) = - \bv(\bk)$. 
Hence, $\Pi_{t,N}$ can be reduced to $\Pi_N$ in the collinear 
low-energy limit only for {\em even}\/ $N$:
\begin{equation} \label{Pi_tN_N}
 \Pi_{t,N}(q_1,\dots,q_N) \to
 2 \prod_{j=1}^N \hat\bq_{j\perp} \cdot \bv(\bk_F) \, 
 \Pi_N(q_1,\dots,q_N) \, .
\end{equation}
For odd $N$, contributions from $\bk$ near $\bk_F$ and $-\bk_F$
contribute with opposite sign and can therefore not be written
in terms of $\Pi_N$. However, the results obtained for even
$N$ suffice to show that the effective action involves 
non-local marginal interaction of arbitrarily high order.

The symmetrized $N$-point loop
\begin{equation} \label{Pi_N^S}
 \Pi_N^S(q_1,\dots,q_N) = \frac{1}{N!} \sum_{P}
 \Pi_N(q_{P1},\dots,q_{PN})
\end{equation}
describes the dynamical $N$-point {\em density correlations}\/ 
of a Fermi gas.
In the following sections we will derive its scaling behavior
in the collinear low-energy limit for arbitrary $N$.

%%% Exact results for N-point density loop %%%%%%%%%%%%%%%%%%%%%%%

\section{Exact formulae for N-point density loop}

Our analysis of the scaling behavior of $\Pi_N$ and $\Pi_N^S$ 
is based on exact expressions derived by Feldman et al. 
\cite{feldman98} and their elaboration by Neumayr and Metzner.
\cite{neumayr98}
They are valid for a parabolic dispersion relation.
In this section we summarize these expressions, assuming
specifically $\eps(\bk) = \bk^2/2$ and $k_F = 1$.
Obviously one may restore an arbitrary mass and $k_F$ at will.
We use the parametrization $I_N(p_1,\dots,p_N)$ with momenta
$p_j$ linearly related to the $q_j$, as described in Sec.~II,
see also Fig.~1.
The following expressions are applicable only for non-collinear 
momenta. Nevertheless, they can be used to study the scaling 
behavior in a limit where they become increasingly collinear
upon reducing $\lam$.

The $N$-point loop can be expressed as a linear combination 
of $3$-point loops with rational coefficients:
\cite{feldman98,neumayr98,neumayr99}
\begin{equation} \label{reduc}
 I_N(p_1,\dots,p_N) = \sum_{1 \leq i < j < k \leq N}
 \left[ \prod_{\nu = 1 \atop \nu \neq i,j,k}^N
 f_{i\nu}(\bd^{ijk}) \right]^{-1}
 I_3(p_i,p_j,p_k) \, ,
\end{equation}
where
\begin{eqnarray} \label{f(d)}
 f_{i\nu}(\bd^{ijk}) &=&
 \frac{1}{2} (\bp_i^2 - \bp_{\nu}^2) + 
 i(p_{i0} - p_{\nu 0})
 \nonumber \\[2mm]
 &+& \left\{ \left[ 
 \frac{1}{2} (\bp_k^2 - \bp_i^2) + 
 i(p_{k0} - p_{i0}) \right]
 \frac{\det(\bp_j - \bp_i,\bp_{\nu} - \bp_i)}
 {\det(\bp_j - \bp_i,\bp_k - \bp_i)} + 
 j \leftrightarrow k \right\} \, .
\end{eqnarray}
The determinants of two momenta are defined as
$\det(\bp,\bp') = \det \left( p_x \, p'_x \atop p_y \, p'_y \right)$.

For the 3-point loop, Feldman et al. \cite{feldman98} have obtained 
the expression
\begin{equation} \label{I_3}
 I_3(p_1,p_2,p_3) = 
 \frac{1}{2\pi i \, \det(\bp_2\!-\!\bp_1,\bp_3\!-\!\bp_1)}
 \sum_{i,j=1 \atop i \neq j}^3 s_{ij} \, t_{ij}
\end{equation}
where $s_{12} = s_{23} = s_{31} = 1$, $s_{21} = s_{32} = s_{13} = -1$,
and
\begin{equation} \label{t_ij}
 t_{ij} = \int_{\gam_{ij}} \frac{dz}{z} \, .
\end{equation}
The contour-integrals are performed along the curves
$\gam_{ij} = \{ w_{ij}(s) | 0 \leq s \leq 1 \} $,
where $w_{ij}(s)$ is the unique (generally complex) root of the 
quadratic equation
\begin{equation}\label{w-eq}
 (\bp_j\!-\!\bp_i)^2 \, z^2 + 
 2 \, \det(\bd\!-\!\bp_i,\bp_j\!-\!\bp_i) \, z +
 (\bd\!-\!\bp_i)^2 = s^2  \, ,
\end{equation}
satisfying the condition
\begin{equation} \label{Imcondition}
 \Im \, \frac{-(p_{jx} - p_{ix}) z + d_y - p_{iy}}
          { (p_{jy} - p_{iy}) z + d_x - p_{ix}} >  0 \, .
\end{equation}
The (complex) two-dimensional vector $\bd = (d_x,d_y)$ is given by
\begin{equation}
 \bd = \frac{1}{\det(\bp_2 - \bp_1,\bp_3 - \bp_1)} \,
 \left[ \frac{1}{2} (\bp_3^2 - \bp_1^2) + i(p_{30} - p_{10}) 
 \right] (\bp_2 - \bp_1)_{\perp} \, + \, 
 p_2 \leftrightarrow p_3 \, .
\end{equation}
The integration path $w_{ij}(s)$ can be written explicitly as
\cite{neumayr98}
\begin{equation} \label{w_ij}
 w_{ij}(s) = \frac{z_{ij}(s) - \bar z_{ij}}{|\bp_j - \bp_i|}
\, .
\end{equation}
Here $z_{ij}(s) = x_{ij}(s) + iy_{ij}(s)$ is a function of 
$s$ with real and imaginary parts given by
\begin{eqnarray}
\label{x_ij}
 x_{ij}(s) &=& \sgn(p_{j0} - p_{i0}) \, \frac{1}{\sqrt{2}} \,
 \left[ \sqrt{[a_{ij}(s)]^2 + (p_{j0} - p_{i0})^2} + 
 a_{ij}(s) \right]^{1/2} \, , \\[1mm]
\label{y_ij}
 y_{ij}(s) &=& - \frac{1}{\sqrt{2}} \,
 \left[ \sqrt{[a_{ij}(s)]^2 + (p_{j0} - p_{i0})^2} - 
 a_{ij}(s) \right]^{1/2} \, ,
\end{eqnarray}
with
\begin{equation} \label{a_ij}
 a_{ij}(s) = s^2 - \frac{1}{4} |\bp_j - \bp_i|^2 +
 \frac{(p_{j0} - p_{i0})^2}{|\bp_j - \bp_i|^2} \, .
\end{equation}
The constant $\bar z_{ij}$ is given by 
$\bar z_{ij} = \bar x_{ij} + i \bar y_{ij} = 
 \bar x_{ijk} + i \bar y_{ijk}$, 
where $k$ completes the index set $\{ i,j \}$ to 
$\{ i,j,k \} = \{ 1,2,3 \}$, and
\begin{eqnarray}
\label{x_ijk}
 \bar x_{ijk} &=& 
 \frac{|\bp_j - \bp_i|}{2 \det(\bp_j - \bp_i,\bp_k - \bp_i)} \,
 (\bp_j - \bp_k) \cdot (\bp_k - \bp_i) \, , \\[2mm]
\label{y_ijk}
 \bar y_{ijk} &=& 
 \frac{(\bp_k - \bp_i)(p_{j0} - p_{k0}) - 
 (\bp_j - \bp_k)(p_{k0} - p_{i0})}
 {\det(\bp_j - \bp_i,\bp_k - \bp_i)} \cdot
 \frac{\bp_j - \bp_i}
 {|\bp_j - \bp_i|} \, .
\end{eqnarray}
One can easily show that $y_{ij}(s)$ increases strictly 
monotonically as a function of $s$, and $x_{ij}(s)$ increases
(decreases) strictly monotonically if $\sgn(p_{j0} - p_{i0}) > 0$
($\sgn(p_{j0} - p_{i0}) < 0$).
The integration path $w_{ij}(s)$ thus has a simple shape.
We finally note the following obvious symmetries under exchange 
of $i$ and $j$:
\begin{equation} \label{sym1ij}
 x_{ji}(s) = - x_{ij}(s) \, \quad y_{ji}(s) = y_{ij}(s) \, ,
\end{equation}
\begin{equation} \label{sym2ij}
 \bar x_{ji} = - \bar x_{ij} \, \quad 
 \bar y_{ji} = - \bar y_{ij} \, .
\end{equation}
%

%%% 3-point loop in collinear low-energy limit %%%%%%%%%%%%%%%%%%%%%

\section{3-point loop in collinear low-energy limit}

We now derive the asymptotic behavior of the 3-point loop 
$I_3(p_1,p_2,p_3)$ in the collinear low-energy limit.
To this end we substitute
$p_{j0} \mapsto \lam^3 p_{j0}$, $p_{jx} \mapsto \lam^2 p_{jx}$,
$p_{jy} \mapsto \lam p_{jy}$ and expand $I_3$ as given by 
Eq.~(\ref{I_3}) in powers of $\lam$.

We first expand the integration path $w_{ij}(s)$, Eq.~(\ref{w_ij}).
For the constants $\bar x_{ijk}$ and $\bar y_{ijk}$ one obtains
\begin{eqnarray}
\label{x_ijk_exp}
 \bar x_{ijk} &=& 
 \frac{|p_{jy} - p_{iy}|}{2 \det(\bp_j - \bp_i,\bp_k - \bp_i)} \,
 (p_{jy} - p_{ky}) (p_{ky} - p_{iy}) + \cO(\lam) \, , \\[2mm]
\label{y_ijk_exp}
 \bar y_{ijk} &=& 
 \lam \frac{(p_{ky} - p_{iy})(p_{j0} - p_{k0}) - 
 (p_{jy} - p_{ky})(p_{k0} - p_{i0})}
 {\det(\bp_j - \bp_i,\bp_k - \bp_i)} \,
 \frac{p_{jy} - p_{iy}}{|p_{jy} - p_{iy}|} + \cO(\lam^2) \, .
\end{eqnarray}
For the functions $x_{ij}(s)$ and $y_{ij}(s)$ one finds
\begin{eqnarray}
\label{x_ij_exp}
 x_{ij}(s) &=& \sgn(p_{j0} - p_{i0}) s + 
 \cO(\lam^2) \, , \\[2mm]
\label{y_ij_exp}
 y_{ij}(s) &=& - \lam^3 \frac{|p_{j0} - p_{i0}|}{2s} + 
 \cO(\lam^5) \, ,
\end{eqnarray}
for $s > 0$, and
\begin{eqnarray}
\label{x0_ij_exp}
 x_{ij}(0) &=& 
 \lam^2 \frac{p_{j0} - p_{i0}}{|p_{jy} - p_{iy}|} + 
 \cO(\lam^3) \, , \\[2mm]
\label{y0_ij_exp}
 y_{ij}(0) &=& - \frac{\lam}{2} |p_{jy} - p_{iy}| + 
 \cO(\lam^2) \, .
\end{eqnarray}
Inserting the expansion of the above auxiliary quantities
into Eq.~(\ref{w_ij}), and splitting the real and imaginary
parts, one obtains
\begin{eqnarray} \label{w_ij_0exp}
 w_{ij}(0) &=& - \frac{1}{\lam} \,
 \frac{(p_{jy} - p_{ky})(p_{ky} - p_{iy})}
 {2\det(\bp_j - \bp_i,\bp_k - \bp_i)} + \cO(1)
 \nonumber \\
 &-& i \left[ \frac{1}{2} + 
 \frac{(p_{ky} - p_{iy})(p_{j0} - p_{k0}) - 
 (p_{jy} - p_{ky})(p_{k0} - p_{i0})}
 {\det(\bp_j - \bp_i,\bp_k - \bp_i)} \,
 \frac{p_{jy} - p_{iy}}{|p_{jy} - p_{iy}|^2} + \cO(\lam)
 \right] \, , \hskip 1cm
\end{eqnarray}
and
\begin{equation} \label{w_ij_exp}
 w_{ij}(s) = w_{ij}(0) + 
 \frac{s}{\lam} \, \frac{\sgn(p_{j0} - p_{i0})}{|p_{jy} - p_{iy}|}
 + \cO(1) + i \left[ \frac{1}{2} + \cO(\lam) \right]
\end{equation}
for $s > 0$.

The value of $t_{ij}$, Eq.~(\ref{t_ij}), is given by the
difference of natural logarithms at the end and the beginning
of the integration path, plus contributions $\pm 2\pi i$ 
for each crossing of the branch cut on the negative real axis 
in the complex plane.
For small $\lam$ one can write $t_{ij}$ in a form where
no case-dependent multiples of $2\pi i$ appear, namely
\cite{thier11}
\begin{equation} \label{t_ij_ln}
 t_{ij} = \left. \ln[u_{ij}(s)] + 
 \ln\left[ 1 + i \frac{y_{ij}(s)}{u_{ij}(s)} \right]
 \right|_0^1 \, ,
\end{equation}
where
\begin{equation} \label{u_ij}
 u_{ij}(s) = x_{ij}(s) - \bar x_{ij} - i \bar y_{ij} \, .
\end{equation}

The sum in Eq.~(\ref{I_3}) can be written as
\begin{equation} \label{sum_ij}
 I'_3 =
 \sum_{(i,j) = (1,2),(2,3),(3,1)} (t_{ij} - t_{ji}) \, .
\end{equation}
Forming the difference $t_{ij} - t_{ji}$, the first terms
from Eq.~(\ref{t_ij_ln}) cancel due to the antisymmetry
of $u_{ij}(s)$ in $i$ and $j$, such that
\begin{equation} \label{t_ij_diff}
 t_{ij} - t_{ji} = \left. 
 \ln\left[ 1 + i \frac{y_{ij}(s)}{u_{ij}(s)} \right] -
 \ln\left[ 1 - i \frac{y_{ij}(s)}{u_{ij}(s)} \right]
 \right|_0^1 \, .
\end{equation}
This is a suitable starting point for an expansion in 
powers of $\lam$, since $y_{ij}(s)/u_{ij}(s)$ is of
order $\lam$ for $s=0$, and of order $\lam^3$ for $s=1$.
Expanding the logarithm yields
\begin{equation} \label{t_ij_exp}
 t_{ij} - t_{ji} = \sum_{n=0}^{\infty} 
 \frac{2}{2n + 1} \, \left\{
 \left[ i \frac{y_{ij}(1)}{u_{ij}(1)} \right]^{2n+1} -
 \left[ i \frac{y_{ij}(0)}{u_{ij}(0)} \right]^{2n+1}
 \right\} \, .
\end{equation}
Inserting $u_{ij}(s)$ from Eq.~(\ref{u_ij}) and expanding
in powers of $\lam$, one obtains
\begin{eqnarray} \label{t_ij_exp2}
 t_{ij} - t_{ji} &=&
 2i \frac{y_{ij}(0)}{\bar x_{ij}} +
 2 \frac{\bar y_{ij} \, y_{ij}(0)}{\bar x_{ij}^2} 
 \nonumber \\
 &+& 2i \left[ \frac{y_{ij}(1)}{x_{ij}(1) - \bar x_{ij}}
 + \frac{x_{ij}(0) \, y_{ij}(0)}{\bar x_{ij}^2} 
 - \frac{\bar y_{ij}^2 \, y_{ij}(0)}{\bar x_{ij}^3}
 - \frac{y_{ij}^3(0)}{3 \bar x_{ij}^3} \right] 
 + \cO(\lam^4) \, ,
\end{eqnarray}
where the first term is of order $\lam$, the second of
order $\lam^2$, and the third one of order $\lam^3$.

In the sum over pairs $(i,j)$, from Eq.~(\ref{sum_ij}), 
contributions of order $\lam$ and $\lam^2$ to the single 
differences $t_{ij} - t_{ji}$ cancel, as do many terms
of order $\lam^3$.
To see this one has to insert expansions of the auxiliary
quantities appearing in Eq.~(\ref{t_ij_exp2}) in powers
of $\lam$, sometimes beyond the order presented in 
Eqs.~(\ref{x_ijk_exp}) - (\ref{y0_ij_exp}).
After a lengthy but straightforward calculation one
obtains \cite{thier11}
\begin{equation} \label{sum_ij_exp}
 I'_3 = 2i \sum_{(i,j) = (1,2), (2,3), (3,1)}
 \frac{y_{ij}(1)}{x_{ij}(1) - \bar x_{ij}} + 
 \cO(\lam^4) \, .
\end{equation}
Expanding $\bar x_{ij}$, $x_{ij}(1)$, and $y_{ij}(1)$
yields
\begin{equation} \label{frac_exp}
 \frac{y_{ij}(1)}{x_{ij}(1) - \bar x_{ij}} =
 - \frac{\lam^3}{2} \frac{D_{ijk} (p_{j0} - p_{i0})}
 {D_{ijk} + \frac{1}{2} \sgn(p_{j0} - p_{i0}) \, 
  \sgn(p_{jy} - p_{iy}) F_{ijk}}
 + \cO(\lam^4) \, ,
\end{equation}
where
\begin{equation} \label{D_ijk}
 D_{ijk} = \det(\bp_j - \bp_k,\bp_i - \bp_k) \, ,
\end{equation}
and
\begin{equation} \label{F_ijk}
 F_{ijk} = 
 (p_{ky} - p_{jy})(p_{jy} - p_{iy})(p_{iy} - p_{ky}) \, .
\end{equation}
Note that $D_{ijk}$ and $F_{ijk}$ are both invariant
under cyclic permutations of $i$,$j$, and $k$.
Inserting Eq.~(\ref{frac_exp}) into Eq.~(\ref{sum_ij_exp}),
and dividing by $2\pi i \, \det(\bp_2 - \bp_1,\bp_3 - \bp_1)$,
we obtain our final result for the collinear low-energy limit
of the 3-point loop
\begin{eqnarray} \label{I_3_exp}
 \Pi_3(q_1,q_2,q_3) &=& I_3(p_1,p_2,p_3)
 \nonumber \\[1mm]
 &=& \frac{1}{2\pi} \sum_{(i,j,k) = (1,2,3) + cyc.}
 \frac{p_{j0} - p_{i0}}
 {D_{123} + \frac{1}{2} \sgn(p_{j0} - p_{i0}) \, 
  \sgn(p_{jy} - p_{iy}) F_{123}} + \cO(\lam) \, . \hskip 1cm
\end{eqnarray}

The 3-point loop is thus generically finite for $\lam \to 0$.
The limit is real and depends on the ratios 
$(p_{j0} - p_{i0})/D_{123}$ and $(p_{j0} - p_{i0})/F_{123}$.
Note that $|F_{123}| = q_{1y} q_{2y} q_{3y}$, while 
$|D_{123}|$ is twice the area of the triangle with corners
$\bp_1$, $\bp_2$, and $\bp_3$, or, equivalently, of the triangle
obtained by attaching the vectors $\bq_1$, $\bq_2$ and 
$\bq_3$ to each other.
$\Pi_3(q_1,q_2,q_3)$ vanishes if frequency variables $q_{j0}$ 
are set to zero before scaling $\bq_j$ to zero, in agreement 
with the result of Hertz and Klenin.\cite{hertz74}
It also vanishes if either 
$\sgn(p_{j0} - p_{i0}) = \sgn(p_{jy} - p_{iy})$ 
for all $(i,j)$ or
$\sgn(p_{j0} - p_{i0}) = - \sgn(p_{jy} - p_{iy})$ 
for all $(i,j)$.
In these cases the contributions to the sum over cyclic
permutations of $(1,2,3)$ in Eq.~(\ref{I_3_exp}) cancel.

The expression on the right hand side of Eq.~(\ref{I_3_exp}) 
is invariant under permutations of $q_1$, $q_2$, and $q_3$, 
so that it also describes the collinear low-energy limit of the 
symmetrized 3-point loop $\Pi_3^S(q_1,q_2,q_3)$.

%%% N-point loop in collinear low-energy limit %%%%%%%%%%%%%%%%%%%%%

\section{N-point loop in collinear low-energy limit}

The reduction formula (\ref{reduc}) relates the $N$-point loop
to a linear combination of 3-point loops.
The coefficients are determined by the quantities 
$f_{i\nu}(\bd^{ijk})$ defined in Eq.~(\ref{f(d)}).
In the collinear low-energy limit, the latter become frequency
independent and real, and they scale as
\begin{eqnarray} \label{f(d)_exp}
 f_{i\nu}(\bd^{ijk}) &=&
 \frac{\lam^2}{2} (p_{iy}^2 - p_{\nu y}^2)
 \nonumber \\[2mm]
 &+& \frac{\lam^2}{2} \left[
 (p_{ky}^2 - p_{iy}^2) 
 \frac{D_{ij\nu}}{D_{ijk}} + 
 j \leftrightarrow k \right] + \cO(\lam^3) \, .
\end{eqnarray}
Inserting this and Eq.~(\ref{I_3_exp}) for $I_3$ into 
Eq.~(\ref{reduc}), one obtains an explicit formula for the
$N$-point loop in the collinear low-energy limit.
It is remarkable that 
$\Pi_N(q_1,\dots,q_N) = I_N(p_1,\dots,p_N)$ is a 
{\em rational}\/ function of all momenta and frequencies
in this limit.
For $N > 3$, it {\em diverges} as
\begin{equation} \label{Pi_N_div}
 \Pi_N \propto \lam^{2(3-N)}
\end{equation}
for $\lam \to 0$.

The degree of divergence of $\Pi_N$ is not reduced 
upon symmetrization, so that the symmetrized $N$-point 
loop also diverges as
\begin{equation} \label{Pi_N^S_div}
 \Pi_N^S \propto \lam^{2(3-N)} \, .
\end{equation}
We have confirmed the absence of significant cancellations
by computing the scaling behavior of $\Pi_N^S$ for various 
choices of $N$ and $q_1,\dots,q_N$.
This result is remarkable since strong and systematic
cancellations have been shown to occur upon symmetrization
when the limit $q_j \to 0$ is taken more conventionally.
In particular, uniformly scaled $N$-point loops
$\Pi_N(\lam q_1,\dots,\lam q_N)$ diverge as $\lam^{2-N}$,
while their symmetrized counterparts
$\Pi_N^S(\lam q_1,\dots,\lam q_N)$ remain finite.
\cite{neumayr98,kopper01}
The arguments establishing the cancellation of divergences
in the uniform small-$q$ limit do not apply in the 
collinear low-energy limit.

The divergence in Eq.~(\ref{Pi_N^S_div}) is the
``worst case scenario'' compatible with simple power 
counting:
In the collinear low-energy limit the integration measure
in the definition of the $N$-point loop scales as
$\lam^6$, while each of the $N$ propagators diverges as
$\lam^{-2}$, such that the loop may diverge as 
$\lam^{6-2N}$ (but not stronger).
What we have shown is that this divergence is neither
reduced by oscillations of the integrand under the 
$k$-integral in Eq.~(\ref{Pi_N}), nor by cancellations 
in the sum over permutations contributing to the 
symmetrized loops.
Note that the power-counting does not change if the bare 
propagator $G_0(k)$ in the $N$-point loops is replaced by 
a propagator $G(k)$ with a self-energy proportional to 
$k_0^{2/3}$, since this interacting propagator also diverges
as $\lam^{-2}$.

%%% Power counting of order parameter interaction %%%%%%%%%%%%%%%%

\section{Power counting of order parameter interaction}

Now that we have determined the scaling behavior of the $N$-point
loop in the collinear low-energy limit, we can assess the relevance
of the $N$-point order parameter interactions in the effective action
$S[\phi]$ by using standard power counting.
To see how the interaction terms evolve compared to the quadratic
part of the action, we rescale the field $\phi$ in such a way that 
the bare critical action $S_0[\phi]$, Eq.~(\ref{S0}), remains 
invariant.
Since the integration measure scales as $\lam^6$, and the inverse
bare propagator $D_0^{-1}(q)$ as $\lam^2$, we have to rescale
the field as $\phi \mapsto \lam^{-4} \phi$.
The $N$-point interaction terms in $S[\phi]$ are composed of $N$ 
energy-momentum integrals, a delta-function for energy-momentum
conservation, the $N$-point loop, and a product of $N$ fields
$\phi(q_1) \dots \phi(q_N)$.
The $N$-point interaction therefore scales as
\begin{equation} \label{S_IN}
 S_{I,N}[\phi] \propto
 \lam^{6N} \lam^{-6} \lam^{2(3-N)} \lam^{-4N} = \lam^0 \, .
\end{equation}
All $N$-point interactions contributing to $S[\phi]$ are thus
{\em marginal}\/ in the collinear low-energy scaling limit.
Hence, the effective order parameter action cannot be truncated
at any finite order. At least such a truncation is not justified
by power counting.
Furthermore, the interaction terms have singular momentum and
energy dependences, which cannot be represented by local
interactions.

Whether the Gaussian fixed point remains stable or not depends
therefore entirely on the behavior of fluctuation contributions.
To get an idea one might compute low order fluctuation
corrections to $S_{I,3}[\phi]$ and $S_{I,4}[\phi]$.
Metlitski and Sachdev \cite{metlitski10nem} have shown that 
the bare Gaussian propagator $D_0(q)$ does not receive qualitative 
modifications up to three-loop order in the coupled 
fermion-boson theory underlying the effective action $S[\phi]$.
\cite{footnote1,footnote2}
Obtaining a general conclusion on the fate of the Gaussian
fixed point seems difficult, however, since the theory has no 
obvious expansion parameter.

Let us also discuss the power counting for a generalized
Gaussian part of the effective action, of the form
$S_0[\phi] = \frac{1}{2} \int_q 
 \big( \chi \bq^{1+\eps} + \gam \frac{|q_0|}{|\bq|} \big) 
 \phi(q) \phi(-q)$, where $\eps \in [0,1]$.
This generalization was introduced by Nayak and Wilczek 
\cite{nayak94} for the sake of a controlled expansion in
$\eps$. 
It was recently used by Mross et al.\cite{mross10} for the 
purpose of defining a managable large-$N_f$ limit of the 
theory, where $N_f$ is the number of fermion flavors.
The case $\eps = 0$ is related to the theory of electrons
in a half-filled Landau level,\cite{halperin93} while
$\eps = 1$ describes the nematic QCP and related systems
as discussed above.
For arbitrary $\eps$, the scaling limit (\ref{colllimit})
has to be generalized to
$q_{j0} \mapsto \lam^{2+\eps} q_{j0}$,
$q_{jx} \mapsto \lam^2 q_{jx}$, and
$q_{jy} \mapsto \lam   q_{jy}$,
corresponding to a dynamical exponent $z = 2 + \eps$.
The limit is still collinear and ``static'' (in the sense
that frequencies scale to zero faster than the modulus
of momenta). 
In light of the results for $\eps = 1$ it is thus likely 
that the degree of divergence of the $N$-point loops is not
reduced by cancellations in this scaling limit for $\eps < 1$,
too, such that it can be estimated by naive power counting.
One then obtains $\Pi_N^S \propto \lam^{5+\eps-2N}$.
Powercounting for the effective action then yields 
$S_{I,N}[\phi] \propto \lam^0$ as in the special case 
$\eps = 1$. The $N$-point interactions are thus marginal
in the collinear low-energy limit, for any $N$ and $\eps$.
The size of $\eps$ does not matter here.

We finally compare to the situation in three dimensions.
Here finite $\phi^{N}$ interactions are even more irrelevant
than in two dimensions, but one may again wonder about the 
special role of momentum transfers tangential to the Fermi
surface in a certain Fermi point (and its antipode), which
are mutually {\em coplanar}.
Considering the coplanar low-energy scaling limit 
$q_0 \mapsto \lam^3 q_0$, $q_x \mapsto \lam^2 q_x$, and 
$q_{y,z} \mapsto \lam q_{y,z}$, one expects a divergence
$\Pi_N^S \propto \lam^{7 - 2N}$ for $N \geq 4$.
These divergences are however not strong enough to make
$S_{I,N}[\phi]$ marginal or relevant.
Power counting yields $S_{I,3}[\phi] \propto \lam^{1/2}$,
and $S_{I,N}[\phi] \propto \lam^{N/2}$ for $N \geq 4$.
The Gaussian fixed point is thus clearly stable in three
dimensions.
%Small-q QCPs with Ising order parameter in 3D are thus HM.

%%% Conclusion %%%%%%%%%%%%%%%%%%%%%%%%%%%%%%%%%%%%%%%%%%%%%%%%%%%

\section{Conclusion}

In summary, we have analyzed the scaling behavior of $N$-point 
interactions between order parameter fluctuations at a nematic 
QCP and for other quantum critical electron systems with a 
scalar order parameter in two dimensions.
The $N$-point interactions are given by symmetrized fermionic
loops with $N$ vertices.
We have shown that these loops exhibit a singular momentum and
energy dependence for all $N \geq 3$, so that they cannot be
represented by a local interaction.
For $N \geq 4$, they diverge in the collinear low-energy limit, 
where energy variables scale to zero faster than momenta, and 
momenta become increasingly collinear.
We have derived explicit expressions for the momentum and 
energy dependences in that limit.
The degree of divergence is not reduced by any cancellations.
From standard power counting one then obtains that all $N$-point 
interactions are marginal.

The effective action is thus dominated by interactions between
fluctuations with collinear momenta, as noted already previously.
\cite{metlitski10nem}
It cannot be truncated at any finite order and none of the
$N$-point terms can be represented by a local interaction.
In particular, approximating the interactions by a local quartic
term, as in Hertz' theory, is inadequate.
The same conclusion can be drawn for the effective action
describing fermions coupled to a $U(1)$ gauge field.

Marginality of all $N$-point order parameter interactions has
already been obtained for a spin density wave QCP in two 
dimensions.\cite{abanov04}
In that case the fluctuation propagator is strongly renormalized 
by these interactions. In particular, anomalous scaling dimensions 
appear.\cite{abanov03,metlitski10sdw}
For the nematic QCP and related theories it is presently unclear
whether the Gaussian fixed point remains stable.
Perturbative calculations have not yet revealed any singular
renormalization of the bare Gaussian propagator.
\cite{kim94,metlitski10nem}

\begin{acknowledgments}
We are grateful to D.~Belitz, A.~Chubukov, M.~Metlitski, 
and S.~Sachdev for valuable discussions,
and to N.~Hasselmann for a critical reading of the manuscript.
\end{acknowledgments}

%%%%%%%%%%%%%%%%%%%%%%%%%%%%%%%%%%%%%%%%%%%%%%%%%%%%%%%%%%%%%%%%%%%%


\begin{thebibliography}{99}
\bibitem{footnote1} The $\bq^2$-term survives to three-loop order
 and the dynamical exponent remains $z=3$. 
 However, in a generalization to an arbitrary number of fermion
 flavors $N_f$, the prefactor of the $\bq^2$-term acquires a dependence 
 on $N_f$ which is very different from the RPA result.
\end{thebibliography}
\end{document}